\documentclass{epl}
\input epsf
\title{
Weakly versus highly nonlinear dynamics 
in 1D systems
}
\author{O.\ Pierre-Louis}
\institute{CNRS/Laboratoire de Spectrom\'etrie Physique,
U.J.Fourier-Grenoble 1,
BP87, F38402 St Martin d'H\`eres, France.}

\begin{document}
\maketitle

\begin{abstract}
We analyze the morphological transition
of a one-dimensional system described by a scalar 
field, where a flat
state looses its stability. This scalar field 
may for example account for the position of a 
crystal growth front, an order parameter,
or a concentration profile.
We  show that two types of dynamics
occur around the transition:
weakly nonlinear dynamics,
or highly nonlinear dynamics. 
The conditions under which highly nonlinear evolution equations 
appear are determined, and their generic form is derived.
Finally,  examples are discussed.
\end{abstract}

In the study of pattern formation,
weakly nonlinear equations play a central role.
By construction, these equations 
catch the main effects of nonlinearities
via a limited number of nonlinear terms added to
a linear equation. This approach has been used 
for a wide variety of physical
systems such as crystal growth \cite{CG},
reaction-diffusion systems \cite{RD},
flame fronts \cite{FF}, or phase separation \cite{variational}. 
Weakly nonlinear equations 
can be derived from a multi-scale analysis
when separation of scales is possible.
This is for example the case in the vicinity
of an instability threshold, where the
system is weakly unstable, 
or in the analysis of amplitude and
phase dynamics of modulated structures\cite{maneville}.
These equations are also obtained from
renormalization techniques\cite{Chen1994}.

Some analysis and attempt of classification
of generic nonlinear equations based on symmetry
or geometry have already been reported in the literature 
\cite{maneville,cross-ho,misbah-geom}.
The most systematic approach up to now
was that of Ref. \cite{misbah-geom}, where
nonlinear equations result from the expansion
in Cartesian coordinates of dynamics
expressed in intrinsic coordinates.
We here present a more general approach based on a multi-scale
analysis.
For the sake of simplicity, we assume that
dynamics is local and that an instability appears 
at long wavelength at the instability threshold. 
From the assumption that the stabilizing or nonlinear
terms do not scale with the small
parameter of the expansion $\epsilon$,
we find that the Benney,
and the sand ripple\cite{csahok-misbah} 
equations are expected in systems
with translational invariance, and 
that the convective Ginzburg-Landau
equation is expected in absence of translational invariance.

Furthermore, our approach 
determines the range of validity of weakly nonlinear expansions
even when $\epsilon$ is present in the stabilizing
or nonlinear terms.
As a central result, we show that the weakly nonlinear
approach breaks down for a large class of front dynamics.
The main property of these systems is to be in the
vicinity of a variational steady-state
(which may be thermodynamic equilibrium).
We obtain highly non-linear
evolution equations in two cases:
(1) Conserved dynamics with
translational invariance.
We provide explicit illustrations of this
case in the context of Molecular Beam Epitaxy.
(2) Dynamics without
translational invariance, but with a
translationally invariant
stabilization process (such as surface tension).
This situation for example applies to
phase separation \cite{variational}, or to amplitude equations
for modulated patterns \cite{maneville}.

Unexpectedly, highly non-linear equations exhibit the ``central" symmetry
in case (1), and the left-right symmetry in case (2),
which are not present in the original physical problem.
Furthermore, a Lyapunov functional may be
found in case (2), although we are in a fully
non-equilibrium situation.

Let us first consider the case where the front obeys
translational invariance. This means that $\partial_th$
does not change when the whole front is translated
via $h\rightarrow h+h_0$, where $h_0$ is a constant.
Then, dynamics does not depend on the position of the front $h(x,t)$,
but only on its derivatives with respect to time $t$ and space $x$.
For a linear analysis, we consider
a small perturbation corresponding to a unique Fourier
mode $h(x,t)=h_{\omega k}\exp(\omega t+ikx)$.
Inserting this relation in 
any front dynamics model then provides
$\omega$ as a specific function of $k$. 
Assuming that dynamics is local, $\Re e[\omega]$
and $\Im m[\omega]$ respectively only involve 
even and odd powers of $k$. 
A long wavelength (i.e. small $k$) expansion then leads to:
\begin{eqnarray}
\Re e[\omega] = L_2 k^2+L_4k^4+o(k^6);
\hspace{0.5 cm}
\Im m[\omega] = L_3k^3+o(k^5)
\label{e:disp_transl_inv}
\end{eqnarray}
From translational invariance, the 
mode $k=0$ is marginally stable, and thus,
there is no constant term in $\Re e[\omega]$.
Moreover, we have performed a Galilean transform
$x\rightarrow x+L_1t$ in order to eliminate
the linear term $L_1k$ in $\Im m[\omega]$.
As $k\rightarrow 0$, 
one has $L_2k^2\gg L_4k^4$.
Therefore, the criterion for an instability to occur 
(i.e. $\Re e[\omega]>0$) at long
wavelength is simply $L_2>0$.
Since we restrict the analysis to instabilities
occurring at long wavelength at the threshold, one should
generically require $L_4<0$.

We now perform a multi-scale
analysis in the vicinity of the instability threshold
based on an expansion with the small parameter $\epsilon\sim L_2$
\footnote{
We could assume that
$L_4\sim\epsilon^{\delta_4}$, as long as
the instability occurs at long wavelength (i.e. $\delta_4<1$). 
Defining $t'=\epsilon^{\delta_4} t$,
$\gamma'=[\gamma+\delta_4(l-1)]/(1-\delta_4)$,
and $\epsilon'=\epsilon^{1-\delta_4}$, we then
have again an equation of the form Eq.(\ref{e:wnl_transl_inv}).
}.
From Eq.(\ref{e:disp_transl_inv}), the unstable
modes are those for which 
$0\leq k \leq k_c=(-L_2/L_4)^{1/2}\sim \epsilon^{1/2}$,
and the most unstable mode is $k_m=k_c/2^{1/2}$.
Therefore, the unstable modes have 
$k\sim\epsilon^{1/2}$ and
$\Re e[\omega]\sim \epsilon^{2}$,
so that the relevant
spatio-temporal scales are
$x\sim k^{-1}\sim \epsilon^{-1/2}$, and 
$t\sim \Re e[\omega]^{-1}\sim\epsilon^{-2}$
\footnote{Although we do not present it here for the sake
of clarity, we should introduce
a propagative time-scale $\sim\epsilon^{-3/2}L_3^{-1}$, 
related to the dispersive term. 
This would not change the results.}.
Using Eq.(\ref{e:disp_transl_inv})
for the linear part, the general form of a weakly
nonlinear equation is defined as:
\begin{eqnarray}
\partial_th=-L_2\partial_{xx}h-L_3\partial_{xxx}h
+L_4\partial_{xxxx}h
+\epsilon^{\gamma}[\partial_x]^n[\partial_t]^l[h]^m
\label{e:wnl_transl_inv}
\end{eqnarray}
In the last term, the brackets mean that we
account at the same time for all terms
containing $n$ spatial and $l$ 
temporal derivatives, and $m$ times $h$ (where $m>1$),
with  arbitrary numerical prefactors. 
Translational invariance
imposes that $n+l\geq m$.
Moreover, we do not know the value of $\gamma$ a priori,
which is a consequence of the specific properties
of the system. 

We first analyse the self consistency of Eq.(\ref{e:wnl_transl_inv}),
using power counting arguments.
Indeed, nonlinear terms and linear terms  relevant 
to stability should be of the same order 
in $\epsilon$.
Stating that $h\sim \epsilon^\alpha$, we then find:
\begin{eqnarray}
\alpha=[2-\gamma-n/2-2l]/(m-1) \, .
\label{e:alpha}
\end{eqnarray}
The weakly nonlinear equation (\ref{e:wnl_transl_inv})
is the result of an expansion of $\partial_th$ --which is 
an unknown function of the derivatives of $h$-- in the limit
$\epsilon\rightarrow 0$. In order to perform this
expansion, we need all the derivatives of $h$
to be smaller than one. Since $h\sim\epsilon^\alpha$ 
and $x\sim\epsilon^{-1/2}$, we need
$\partial^{n_t}_t\partial^{n_x}_xh\sim\epsilon^{2n_t+n_x/2+\alpha}\ll 1$. 
A sufficient condition
is that the largest derivative $\partial_xh$ is smaller than one.
This reads: $\partial_xh \sim\epsilon^{\alpha+1/2} \ll 1$. 
We shall thus require that $\alpha>-1/2$.
This condition allows the
usual gradient expansion.
Combining the different constraints,
we find the condition:
\begin{eqnarray}
2\gamma+3(l-1)<m-n-l\leq 0 \, ,
\label{e:constr_nm}
\end{eqnarray}
which determines all possible
terms which may intervene in a weakly nonlinear
equation.

We have now determined the conditions (\ref{e:constr_nm}) under which
a weakly nonlinear expansion is well defined.
We shall now deal with the more subtle question
of the self-consistency of the dynamics
resulting from a weakly nonlinear equation.
The central question is the interplay between 
dominant and subdominant terms. 
We define the rescaled variables
$\bar h=\epsilon^{-\alpha}h$,
$\bar x=\epsilon^{1/2}x$, and $\bar t=\epsilon^2t$.
Let us then consider two nonlinear terms $N_1$ and $N_2$
(with their rescaled forms $\bar N_1$ and $\bar N_2$)
and the related values of $\alpha$ from Eq.(\ref{e:alpha})
such that $\alpha_1>\alpha_2$
\footnote{$N_i$, with $i=1,2$ more generally describe a sum of
nonlinear terms having the same value of $\alpha_i$.}.
Assuming that $\alpha=\alpha_1$,
one finds that
\begin{eqnarray}
N_2/N_1\sim \epsilon^{(m_2-1)(\alpha_1-\alpha_2)} (\bar N_2/\bar N_1)
\end{eqnarray}
--we recall that $m_2>1$.
Depending on the dynamics in presence of 
$N_1$ in Eq.(\ref{e:wnl_transl_inv}), 
three cases may be observed: (i) The solution $\bar h(\bar x,\bar t)$ 
(or its derivatives
for a system with translational invariance)
is bounded. Then $\bar N_i$ are bounded.
It follows that $N_2\ll N_1$.
(ii) $\bar h(\bar x,\bar t)$ scales with time, i.e.
the typical scales of the patterns evolve as
$\bar h\sim {\bar t}^{a_h}$ and $\bar x\sim {\bar t}^{a_x}$. Then
$\bar N_i\sim {\bar t}^{\kappa_i}$, and 
$N_2/N_1\sim \epsilon^{(m_2-1)(\alpha_1-\alpha_2)}
{\bar t}^{\kappa_2-\kappa_1}$.
Therefore, if $\kappa_1\geq\kappa_2$, we have
$N_1\gg N_2$ at all times. If 
$\kappa_1<\kappa_2$, then $N_2\sim N_1$ after a time 
$t_c\sim t_L\epsilon^{-(m_2-1)(\alpha_1-\alpha_2)}$,
where $t_L$ is the time for appearance of the instability from the linear analysis.
As $\epsilon\rightarrow 0$, one finds 
that $t_c\gg t_L$. 
(iii) $\bar h(\bar x,\bar t)$ 
exhibits an exponential --or 
even faster-- increase of the amplitude (including
the case of singularities in finite time). Then $N_2$
may become of the same order as $N_1$ 
after a time $t_c\sim t_L$
(with possible logarithmic corrections).

From an inspection of the three cases mentioned
above, we see that the only self-consistent
choice in order to describe the dynamics
at timescales that are much longer than $t_L$
is to find, if it exists, the nonlinearity 
with the biggest value of $\alpha$.
The dynamics may then be: (i) bounded, and
the time $t_c$ under which the weakly nonlinear
equation is valid is infinite; (ii)
power-law in time. The time $t_c$ is then
either infinite, or increasing
as $\epsilon$ to some negative power when $\epsilon\rightarrow 0$.
But in the case (iii), where the growth of the
amplitude is exponential or faster, the weakly
nonlinear equation does not provide a satisfactory
description of the dynamics in the nonlinear regime.

We conclude that a nonlinear analysis in the regime where
$\epsilon$ is small  requires
an expansion of the form (\ref{e:wnl_transl_inv}), but also
relies on the knowledge of the dynamics of (\ref{e:wnl_transl_inv})
with the dominant nonlinearities.
There is to our knowledge no general analytical
tool which could systematically determine
whether the nonlinear dynamics belongs to 
one of these 3 classes of dynamics. Therefore, a numerical solution
is in general needed.

Let us now consider some precise examples.
In general, the dominant contribution
depends on the precise values of $\alpha$,
which themselves depends on 
$\gamma$. Therefore, the specific
physical ingredients of the system
will determine the most relevant terms.
Nevertheless, we shall first consider the 
simplified case where $\gamma=0$. 
Relations (\ref{e:constr_nm}) then
readily show that $l=0$. From Eq.(\ref{e:constr_nm}),
we are now able list all possible nonlinearities
in a weakly nonlinear expansion:
$(\partial_xh)^m;
\partial_x(\partial_xh)^m; \partial_{xx}(\partial_xh)^m; 
(\partial_xh)^{m-2}(\partial_{xx}h)^2$, where  $m>1$. 
Using this result we find that
the expected equation is the Benney equation:
\begin{eqnarray}
\partial_th=-\partial_{xx}h+\beta\partial_{xxx}h-\partial_{xxxx}h
+(\partial_xh)^2
\label{e:benney}
\end{eqnarray}
because it leads to the largest possible value
of $\alpha$, which is $\alpha=1$,
and because the solution of Eq.(\ref{e:benney})
leads to a saturation of the amplitude. 
The variables $x,t,h$ have been normalized in
Eq.(\ref{e:benney}) so that only one constant remains.
Eq.(\ref{e:benney}) is non-variational and exhibits
order or chaos when
the parameter $\beta$ is larger or smaller
than one respectively. In the chaotic limit $\beta=0$, it
is called the Kuramoto-Sivashinsky equation.
The Benney and Kuramoto-Sivashinsky equations
have been derived from multi-scale analysis in many
physical situations, such as 
flame fronts \cite{FF}, crystal step meandering
\cite{CG} and bunching \cite{uwaha-sato-misbah-opl}, 
or ion sputtering \cite{sputter}. 

In Molecular Beam Epitaxy (MBE),
atoms land on the surface but can usually not
go back to the atmosphere. 
One then has from mass conservation:
\begin{eqnarray}
\partial_th=F-\partial_xj
\label{e:cons_law}
\end{eqnarray}
where $F$ is the incoming flux, and $j$ is a 
mass flux along the surface. The constant term is
eliminated by the transformation $h\rightarrow h+Ft$.
Many other systems obey a conservation law,
such as wave dynamics in thin liquid layers \cite{whitham},
or sand ripple formation\cite{csahok-misbah}.

Once again, we shall first assume that $\gamma=0$.
From Eqs.(\ref{e:cons_law}) and (\ref{e:constr_nm}), 
the only nonlinear terms which are allowed are:
$\partial_x(\partial_xh)^m;
\partial_{xx}(\partial_xh)^m$.
Following the same line as above, we find that
the first dominant nonlinear term is $\partial_x(\partial_xh)^2$.
But this term  leads to an increase
of the amplitude which is faster than power-law\cite{csahok-misbah}.
This nonlinearity can be absent if it is proportional
to $\epsilon^\gamma$, with $\gamma>1/2$,
but also if some symmetry (such as $h\rightarrow-h$,
or $x\rightarrow-x$) is imposed.
We then find:
\begin{eqnarray}
\partial_th=\partial_x\Bigl[- \partial_xh
+\beta \partial_{xx}h-\partial_{xxx}h
+c_3(\partial_xh)^3+c_2\partial_x(\partial_xh)^2
\Bigr]
\label{e:CM}
\end{eqnarray}
which corresponds to $\alpha=0$.
In Eq.(\ref{e:CM}), $t$, $x$, and $h$ have been normalized,
and $c_i$ are constants.
This equation was first obtained 
by Csahok {\it et al}\cite{csahok-misbah}. 
When $\beta=0$ and $c_2=0$
(e.g. when dynamics is variational,
or when it exhibits the $(h,x)\rightarrow(-h,-x)$
symmetry), and when $c_3<0$, 
one recovers the Cahn-Hilliard \cite{variational}
equation with a double well potential, which is known to lead to
logarithmic coarsening\cite{otha}. 
In this case, the amplitude $\partial_xh$ remains
finite at all times. Therefore, the expansion is self consistent,
and higher order terms are negligible.
As shown in Ref.\cite{csahok-misbah}, this
behavior seems to persist
when $\beta$ and $c_2$ are both non-zero.
But when $c_3=0$ and $c_2\neq 0$,
power law coarsening is found, with
the wavelength $\sim t^{1/2}$
and the amplitude $\sim t^{3/2}$\cite{gillet}.
When $c_3>0$, the dynamics
leads to a local blow up of the slope $\partial_xh$.
Eq.(\ref{e:CM}) describes a wide variety of systems, such as
phase separation \cite{variational}, sand ripple
formation \cite{csahok-misbah}, 
and step bunching \cite{gillet}.

Up to this point, we have found that 
weakly nonlinear expansions are generically
obtained from multi-scale analysis. 
This result was based on the assumption that
$\gamma=0$.  Nevertheless, 
when $\gamma\neq 0$, the set of nonlinear terms which must
be kept can change.
For example, in Ref.\cite{gillet},
the step bunching instability on a vicinal
surface is studied under growth and mobile
atom migration. In this case, 
$\epsilon$ is proportional to the growth rate
for some given value of the migration rate. 
The terms $\partial_x(\partial_xh)^2$ and 
$\partial_x(\partial_xh)^3$ are
absent in Eq.(\ref{e:CM}), because they are multiplied
by $\epsilon^\gamma$ with $\gamma=1$.
A more drastic consequence of a non-vanishing $\gamma$
is the possible break-down of the weakly nonlinear expansion.
In the following, we indeed show that conserved dynamics, 
in the vicinity of thermodynamic equilibrium,
forbids any weakly nonlinear expansion.

Let us first consider the conserved model:
$\partial_th=\partial_x
[M\partial_x(\delta {\cal F}/ \delta h)]$ ,
where $\delta$ denotes the functional derivative, and 
${\cal F}=\int dx \phi$.
$M>0$ and $\phi$ are function of the spatial derivatives of $h$.
${\cal F}$, which plays the role of an energy, and is a Lyapunov functional
(i.e $\partial_t{\cal F}<0$).
Assuming that ${\cal F}$ is minimum for a 
straight front ($\partial_xh=0$), 
any initial perturbation would decay.
This model may for example account for 
relaxation towards thermodynamic equilibrium.
When it is driven by a small non-equilibrium force $f>0$
(not breaking mass conservation), the 
system will respond by
an additional flux $J$. To leading order
in $f$, the new dynamics reads:
\begin{eqnarray}
\partial_th=\partial_x\left[M\partial_x{\delta {\cal F} \over 
\delta h}
+fJ \right]+o(f^2) \, ,
\label{e:wooe}
\end{eqnarray}
where $J$ is a function of the spatial and temporal
derivatives of $h$.
Linearizing Eq.(\ref{e:wooe}), we find:
\begin{eqnarray}
\partial_th=fJ_1\partial_{xx}h+fJ_2\partial_{xxx}h
-(M_0-fJ_3)\partial_{xxxx}h+...
\label{e:lin_cons}
\end{eqnarray}
where $J_1=\partial_{\partial_{x}h}J$,
$J_2=\partial_{\partial_{xx}h}J$,
$J_3=\partial_{\partial_{xxx}h}J$, and
$M_0=M\partial_{\partial_xh\partial_xh}\phi$
in the steady-state configuration.
From the analogy between
Eqs.(\ref{e:lin_cons}) and (\ref{e:disp_transl_inv}),
we conclude that an instability occurs if $J_1<0$,
and that $\epsilon\sim f$. 
From the last term of Eq.(\ref{e:lin_cons})
one then finds that $L_4$ does not scale with $\epsilon$.
We now look for possible weakly nonlinear terms.
For the first term in the brackets
of Eq.(\ref{e:wooe}), $l=0$, $\gamma=0$,
and $n\geq m+3$, which is in contradiction
with (\ref{e:constr_nm}). If a term comes from the
non-equilibrium contribution $J$, one 
has $\gamma=1$, and $n\geq m+1$,
which is again in contradiction with (\ref{e:constr_nm}).
Finally higher order terms have $\gamma\geq 2$,
which is also in contradiction with (\ref{e:constr_nm}).
Therefore, no nonlinear term satisfies (\ref{e:constr_nm}),
and the small gradient constraint 
($\partial_xh\ll 1$) has to be waived.
Thus, we choose $\alpha=-1/2$, and
since $\partial_xh\sim 1$, the full nonlinear dependence
of $M$, $\delta{\cal F}/\delta h$, and $J$ on $\partial_xh$
must be kept, while terms 
such as $\partial_{xx}h$ or $\partial_th$ are negligible.  
Therefore, $J\rightarrow A$, $M\rightarrow B$,
and $\delta{\cal F}/\delta h\rightarrow \partial_xC$, 
where $A$, $B$ and $C$ are functions of
$\partial_xh$ only.
To leading order in $\epsilon$,
Eq.(\ref{e:wooe}) then takes the highly nonlinear form:
\footnote{A more general form for conserved dynamics
when $\alpha=-1/2$
is $\partial_th=\partial_x[B_3\partial_{xx}m
+B_2(\partial_{x}m)^2+B_1\partial_{x}m+
\epsilon B_0m ]$, where $B_i$
are function of $m=\partial_xh$ such that
$B_i\rightarrow$ constant when $m\rightarrow 0$.
}
\begin{eqnarray}
\partial_th=\partial_x\left[B\partial_{xx}C
+\epsilon A \right]
\label{e:hnl_cons_evol}
\end{eqnarray}
Unexpectedly, Eq.(\ref{e:hnl_cons_evol}) is invariant under
the ``central" symmetry $(h,x)\rightarrow(-h,-x)$,
although the starting point Eq.(\ref{e:wooe}) is not.
We shall notice that a formal expansion of $A$, $B$, and
$C$ with respect to $\partial_xh$ in Eq.(\ref{e:hnl_cons_evol})
leads to an equation of the form (\ref{e:wnl_transl_inv})
with an infinite number of nonlinear terms, which all have the
same value of $\alpha$. Therefore, highly nonlinear equations
can be considered as a special case of (\ref{e:wnl_transl_inv}),
and the classification
of the dynamics into the 3 classes, bounded, power-law, and
exponential, is still valid.

An equation of the form (\ref{e:hnl_cons_evol}) was first
derived from a multi-scale analysis of
crystal step meandering during MBE \cite{PierreLouis1999}.
Although Eq.(\ref{e:hnl_cons_evol}) is not variational (except in some
special cases, e.g. when $B\partial_xh=A$ or $C'=B$),
the simple structure of its steady-states
allows one to analyze in details its coarsening
dynamics \cite{politi-misbah} as found in the
case of step meandering from several 
recent works\cite{PierreLouis1999,Paulin2001,Danker2003}.
Eq.(\ref{e:hnl_cons_evol}) was also introduced
as a model for mound formation during MBE\cite{Politi}. 
Our analysis now provides a frame to understand 
its origin.

We now turn to the dynamics of a front without
translational invariance. This situation occurs
when a front, such as the free surface 
of an thin adsorbate, 
is subject to an external field 
which is a function of $h$
(due for example to 
the substrate).
It may also account for unstable concentration
profiles in reaction
diffusion systems \cite{RD}, or for phase
separating systems \cite{variational}.
Since we use a similar approach to that presented above,
we will be more concise here. We shall  assume that there
exist a flat steady state $h=0$.
At long wavelength,
\begin{eqnarray}
\Re e[\omega]\approx L_0+L_2k^2.
\end{eqnarray}
The instability now occurs if $L_0>0$, and we expect $L_2<0$.
We therefore choose $\epsilon\sim L_0$ and the relevant 
spatio-temporal scales are:
$x\sim\epsilon^{-1/2}$
and $t\sim \epsilon^{-1}$. As in the previous case, the term $L_1k$,
which appears in $\Im m[\omega]$, can be
cancelled with the help of a Galilean transform.
The contribution 
$\Im m[\omega]\approx L_3k^3\sim \epsilon^{3/2}$ 
is negligible when $L_3$ does not scale with 
a negative power of $\epsilon$.
The general form of a weakly nonlinear equation is now:
\begin{eqnarray}
\partial_th=L_0 h-L_2\partial_{xx}h
+\epsilon^\gamma[\partial_x]^n[\partial_t]^l[h]^m
\label{e:lin_non_ti}
\end{eqnarray}
with $m>1$.
The nonlinear term will be
of the same order as the linear terms if:
$\alpha=(1-n/2-\gamma-l)/(m-1)$.
The condition for having weakly
nonlinear dynamics is $h\ll 1$,
which implies that $\alpha>0$.
This leads to some restriction,
namely: 
\begin{eqnarray}
\gamma<1-n/2-l.
\end{eqnarray}

Once again, we assume that $\gamma=0$.
Then $l=0$, and the only possible nonlinear terms are:
$h^m ;\partial_x(h^m)$.
The generic first contribution is $h^2$.
It leads to the Fisher-Kolmogorov equation,
which has been extensively used for its traveling
wave solutions \cite{fisher}.
But here, we must start from a front with zero average
height, in which case $h$ locally diverges in finite time.
As before, the presence of a prefactor $\epsilon^\gamma$,
with $\gamma>1/2$, or the existence of symmetries
such as $h\rightarrow-h$ or $(h,x)\rightarrow(-h,-x)$
may forbid this term.
One then finds the convective Ginzburg-Landau equation
\begin{eqnarray}
\partial_th= h+\partial_{xx}h+\sigma h^3+\mu h\partial_xh
\label{e:nc_wnl}
\end{eqnarray}
where $x$ $h$, and $t$ are normalized.
When $\sigma>0$, $h$ again locally diverges in finite
time.
We therefore focus on the case where $\sigma<0$.
Although the Time Dependent Ginzburg-Landau equation 
($\mu=0$Ã) is well known
for phase separating systems\cite{variational} and
as a generic amplitude equation\cite{maneville}, 
we are not aware of previous work incorporating 
the convective term proportional to $\mu$.
This term breaks the
variational character of the dynamics
and the $x\rightarrow-x$ symmetry.
Since there exist two asymmetric kink solutions 
for any value of $\mu$, we do not expect 
qualitative changes in the asymptotic dynamics
of Eq.(\ref{e:nc_wnl}) when $\mu$ varies.
Hence, we conjecture that the known result of logarithmic
coarsening of Eq.(\ref{e:nc_wnl}) for $\mu=0$ \cite{otha}
extends to arbitrary $\mu$. At short times, when $\mu$ is large enough,
Eq.(\ref{e:nc_wnl}) shares similarities with Burgers' equation,
and its solution exhibits shocks.

Let us now consider dynamics having a stable steady state
$h=0$, and a Lyapunov functional ${\cal F}=\int dx \phi$.
The simplest dynamics which has this property
(usually referred to as model A \cite{variational}), is:
$\partial_th=-\Gamma (\delta {\cal F}/\delta h)$,
where $\Gamma$ is a function of $h$ and its spatial derivatives.
In presence of a small destabilizing
non-equilibrium force $f$, we have:
\begin{eqnarray}
\partial_th=-\Gamma {\delta {\cal F} \over \delta h}
+fK + o(f^2) \, ,
\label{e:phys_nti}
\end{eqnarray}
where $K$ depends on $h$ and its spatial and temporal derivatives.
If ${\cal F}$ is translationally invariant,
then $\phi$ is a function of $\partial_xh$ and its
spatial derivatives only. 
Such a situation may be found in
phase separation \cite{variational}, where 
a gradient energy $\sim\int dx  (\partial_xh)^2$
is used, or in the non-conserved dynamics of a 
thin film stabilized by surface tension.
Linearizing Eq.(\ref{e:phys_nti}), 
we then find that $\epsilon\sim f$.
Once again, an inspection of Eq.(\ref{e:phys_nti})
proves that the dynamics is highly nonlinear (i.e. $\alpha=0$).
To leading order in $\epsilon$, the evolution equation
reads:
\footnote{A more general form when $\alpha=0$ is
$\partial_th=P_2\partial_{xx}h+P_1(\partial_xh)^2+
\epsilon P_0 h$, where $P_i$ are functions of
$h$ with $P_i\rightarrow$ constant
when $h\rightarrow 0$. As opposed to Eq.(\ref{e:hnl_nc}),
this equation does not have a Lyapunov functional.}
\begin{eqnarray}
\partial_th=P\partial_{xx}h+\epsilon Q \, ,
\label{e:hnl_nc}
\end{eqnarray}
where $P$ and $Q$ are functions of
$h$ only. We shall notice here
two unexpected and strongly restrictive properties
of this equation: (i) it has the $x\rightarrow-x$
symmetry, although the dynamics
of the original problem does not necessarily have it.
(ii) Although the front is not at equilibrium,
Eq.(\ref{e:hnl_nc}) exhibits a Lyapunov
functional 
\begin{eqnarray}
{\cal U}=\int dx [(\partial_xh)^2/2+\epsilon R],
\end{eqnarray}
where $R$ is a function of $h$ defined by
the relation $R'=Q/P$. Indeed one can write
$\partial_th=-P \delta{\cal U}/\delta h$,
and $\partial_t{\cal U}\leq 0$.
Since Eq.(\ref{e:hnl_nc}) is variational, one
can use the concepts of dynamical scaling
developed for the study of phase transitions \cite{variational}
to study the coarsening.
One should once again rely on the
structure of the steady-states in order 
to analyze the dynamics \cite{politi-misbah}.

To conclude, we have presented
two scenarios for the destabilization of
a 1D system: weakly nonlinear, and highly
nonlinear dynamics.  During weakly nonlinear
dynamics, the evolution of the front 
morphology is described by an nonlinear expansion
at small amplitudes. The amplitude then
tends to zero as one gets closer to
the threshold. During highly nonlinear
dynamics, nonlinearities come into play
only when the amplitude becomes finite.
A small amplitude
expansion is then not justified anymore.  
We have shown that the 
form of the evolution equation can 
still be found in this case.

\end{document}